\documentclass[sigconf, nonacm, pdfa]{acmart}

\newcommand\vldbdoi{XX.XX/XXX.XX}
\newcommand\vldbpages{XXX-XXX}
\newcommand\vldbvolume{18}
\newcommand\vldbissue{12}
\newcommand\vldbyear{2025}

\newcommand\vldbauthors{\authors}
\newcommand\vldbtitle{\shorttitle} 
\newcommand\vldbavailabilityurl{\url{https://github.com/DataManagementLab/demo-explain-lcms}}
\newcommand\vldbpagestyle{empty} 

\usepackage{graphicx}
\usepackage{xurl}
\usepackage{enumitem}
\usepackage{acronym}
\usepackage{multirow}
\usepackage{xspace}
\usepackage{cleveref}
\usepackage[htt]{hyphenat}
\usepackage{booktabs}
\usepackage{graphics}
\usepackage{multicol}
\usepackage{tabularx}
\usepackage[most]{tcolorbox}
\usepackage{xcolor}
\usepackage{tikz}
\usepackage{graphicx}
\usepackage[a-2a,mathxmp]{pdfx}[2018/12/22]
\usetikzlibrary{shapes, arrows, positioning}

\definecolor{operator}{HTML}{E8E8E8}
\definecolor{predicate}{HTML}{FBE3D6}
\definecolor{column}{HTML}{D9F2D0}

\newcommand{\lcm}{\ac{lcm}\xspace}
\newcommand{\lcms}{\acp{lcm}\xspace}
\newcommand{\gnn}{\ac{gnn}\xspace}
\newcommand{\gnns}{\acp{gnn}\xspace}
\newcommand*\circles[1]{\raisebox{.5pt}{\textcircled{\raisebox{-.8pt}{\textsf{#1}}}}}
\newcommand{\sensitivity}{\texttt{SensitivityAnalysis}\xspace}
\newcommand{\gnnexplainer}{\texttt{GNNExplainer}\xspace}
\newcommand{\guidedbackprop}{\texttt{GuidedBackprop}\xspace}
\newcommand{\diffmask}{\texttt{DiffMask}\xspace}
\newcommand{\opbox}[2]{\tcbox[box align=base, nobeforeafter, colback=#1, colframe=black, size=small, left=0pt, right=0pt, boxsep=0.5pt]{\textsf{#2}}\xspace}
\newcommand{\aggregate}{\opbox{operator}{Aggregate}}
\newcommand{\hashjoin}{\opbox{operator}{Hash Join}}

\newcommand{\scanb}{\opbox{operator}{Seq Scan 2}}
\newcommand{\predicate}{\opbox{predicate}{$>=$}}
\newcommand{\column}{\opbox{column}{prod\_year}}
\newcommand{\zeroshot}{\texttt{Zero-Shot}\xspace}

\acrodef{lcm}[LCM]{\emph{Learned Cost Model}}
\acrodef{gnn}[GNN]{\emph{Graph Neural Network}}

\author{Roman Heinrich}
\orcid{0000-0001-7321-9562}
\affiliation{\institution{TU Darmstadt \& DFKI}\country{}}
\email{roman.heinrich@dfki.de}

\author{Oleksandr Havrylov}
\affiliation{\institution{TU Darmstadt}  \country{} }
\email{sashka.havr@gmail.com}

\author{Manisha Luthra}
\orcid{0000-0002-7152-8959}
\affiliation{\institution{TU Darmstadt \& DFKI}\country{}}
\email{manisha.luthra@dfki.de}
 
\author{Johannes Wehrstein}
\orcid{0000-0002-7152-8959}
\affiliation{\institution{TU Darmstadt}\country{}}
\email{johannes.wehrstein@tu-darmstadt.de}

\author{Carsten Binnig}
\orcid{0000-0002-2744-7836}
\affiliation{\institution{TU Darmstadt \& DFKI}\country{}}
\email{carsten.binnig@tu-darmstadt.de}

\title{Opening The Black-Box: Explaining Learned Cost Models For Databases}
\begin{document}
\begin{abstract}
\lcms have shown superior results over traditional database cost models as they can significantly improve the accuracy of cost predictions.
However, \lcms still fail for some query plans, as prediction errors can be large in the tail.
Unfortunately, recent \lcms are based on complex deep neural models, and thus, there is no easy way to understand where this accuracy drop is rooted, which critically prevents systematic troubleshooting.
In this demo paper, we present the very first approach for opening the black box by bringing AI explainability approaches to \lcms.
As a core contribution, we developed new explanation techniques that extend existing methods that are available for the general explainability of AI models and adapt them significantly to be usable for \lcms. 
In our demo, we provide an interactive tool to showcase how explainability for \lcms works. 
We believe this is a first step for making \lcms debuggable and thus paving the road for new approaches for systematically fixing problems in \lcms.
\end{abstract}
\maketitle

\pagestyle{\vldbpagestyle}
\begingroup\small\noindent\raggedright\textbf{PVLDB Reference Format:}\\
\vldbauthors. \vldbtitle. PVLDB, \vldbvolume(\vldbissue): \vldbpages, \vldbyear.\\
\href{https://doi.org/\vldbdoi}{doi:\vldbdoi}
\endgroup
\begingroup
\renewcommand\thefootnote{}\footnote{\noindent
This work is licensed under the Creative Commons BY-NC-ND 4.0 International License. Visit \url{https://creativecommons.org/licenses/by-nc-nd/4.0/} to view a copy of this license. For any use beyond those covered by this license, obtain permission by emailing \href{mailto:info@vldb.org}{info@vldb.org}. Copyright is held by the owner/author(s). Publication rights licensed to the VLDB Endowment. \\
\raggedright Proceedings of the VLDB Endowment, Vol. \vldbvolume, No. \vldbissue\ %
ISSN 2150-8097. \\
\href{https://doi.org/\vldbdoi}{doi:\vldbdoi} \\
}\addtocounter{footnote}{-1}\endgroup

\ifdefempty{\vldbavailabilityurl}{}{
\vspace{.3cm}
\begingroup\small\noindent\raggedright\textbf{PVLDB Artifact Availability:}\\
The source code, data, and/or other artifacts have been made available at \url{https://github.com/DataManagementLab/demo-explain-lcms}.
\endgroup
}

\begin{figure}
    \centering
    \includegraphics[width=0.78\linewidth]{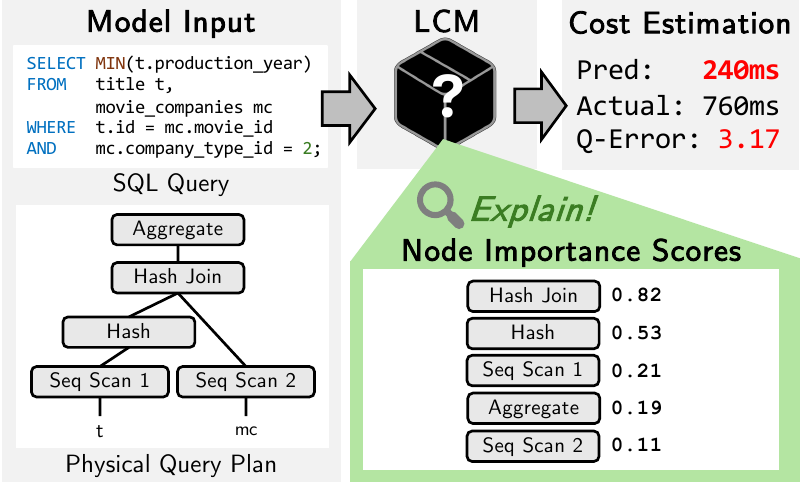} 
    \vspace{-1ex}
    \caption{\lcms predict query execution costs, but they can be off in their predictions. 
    Unfortunately, with current \lcms, it is unclear how they came to their prediction. 
    We leverage explanation algorithms to provide information on which nodes in a query plan contributed most significantly to the runtime prediction, which allows users to understand the model's internal behavior better and contrast it with the actual runtime per operator to identify potential problems of the model as we show in \Cref{fig:detailed_plot}.}
    \label{fig:motivating_plot}
    \vspace{-3ex}
\end{figure}

\section{Introduction}
\indent\noindent{}\textbf{The Potential and Risk of Learned Cost Models.}
\acfp{lcm} for databases were proposed to predict query execution costs \cite{marcus2019, kipf2019, hilprecht2022}. 
Overall, \lcms demonstrate high prediction accuracy, typically outperforming classical approaches \cite{heinrich2025}.
However, \lcms can still produce large errors for individual queries, which is reflected by high tail prediction errors \cite{hilprecht2022}.
Critically, such bad estimates of \lcms lead to sub-optimal plan selections during query optimization, which in turn cause prolonged execution runtimes and thus reduced performance \cite{heinrich2025}.

\noindent{}\textbf{Black-Box Behavior of \lcms.}
Unfortunately, many recent \lcms that provide high precision are based on deep learning models. 
Thus, the predictions are produced by models that are a black-box to database administrators, making it hard to understand their prediction behavior systematically.
In fact, if \lcms provide erroneous predictions, it is unclear \textit{why} they mispredicted the costs for a given plan.
Consequently, a critical downside of \lcms is that we cannot debug them, i.e., trace why they mispredict the costs of a given query plan, which critically prevents systematic troubleshooting.

\noindent{}\textbf{Making \lcms Debuggable.}
In this demo paper, we aim to open the black-box of \lcms and achieve debuggability by making their predictions \textit{explainable}.
The reason is that the knowledge learned by \lcms is hidden in their model parameters. 
Thus, their predictions are inexplicable.
To this end, we propose a novel approach to explain the internalized, learned knowledge of \lcms.
The main idea is that we apply ideas of node importance for \lcms. 
Overall, we believe our approach thus helps to improve the design of future approaches, as it enables the identification of biases in the models, learning strategies, or training data. 

\noindent{}\textbf{An Example.}
We illustrate our approach on a high level in \Cref{fig:motivating_plot}.
The task of a \lcm (middle) is to estimate query execution costs.
Thus, it takes a SQL query as input (left), typically in the form of the physical plan, as this provides information about the planned execution.
The \lcm predicts a query runtime (right) for the given plan, which ideally is close to the actual predicted costs, reflected by a small prediction error.
However, in the example, the \lcm predicts a very low runtime of \texttt{240ms} that significantly ($\approx 3.17\times)$ deviates from the actual runtime of \texttt{760ms}, which is clearly an undesirable behavior.
Given such a query plan, our demo allows to \textit{explain} the prediction of the \lcm.
In particular, we leverage and adapt various explainer algorithms, like \gnnexplainer \cite{ying2019}, to indicate how important every node of the query plan was for the prediction.
In our example, the explainer returns a high importance score of \texttt{0.82} for the \hashjoin, indicating that this node was highly important during the prediction.
Overall, this correlates with the actual importance, as the \hashjoin had the highest runtime during the execution.
However, the explainer assigned a feature importance of \texttt{0.19} for the \aggregate, which is higher than the score for \scanb of \texttt{0.11}.
This is unexpected, as it is not reflected in the actual execution runtimes in this query plan and indicates that the \lcm is not yet able to rank the costs of both operators correctly.

\noindent{}\textbf{Our Contribution: Explainability for \lcms.}
At the core, our novel approach identifies which operators of a query plan were most important for the runtime prediction according to the \lcm.
To achieve this, we significantly adapted existing explainability approaches as follows (cf. \Cref{sec:algos}):
First, we adapted explainers for regression, which is required for cost estimation, as existing explainers typically only support classification.
Second, we introduce a new masking strategy for query operators, which allows us to determine which operators have high relevance for predictions.
Third, we propose novel metrics to explain \lcms, such as the correlation between the actual runtime and the node importance scores (cf. \Cref{subsec:metrics}).
We also integrated our approach into an interactive tool (cf. \Cref{sec:demo}) that allows an interactive comparison between the internalized model knowledge and the actual dependencies of a query plan.

\section{Explaining LCMs}
In the following, we first provide an overview of \lcms to provide the required context before we describe our methodology and the proposed explanation metrics.

\subsection{Learned Cost Models}
The core idea of a \lcm is to learn query execution costs from previously executed query traces.
As neural networks can learn arbitrarily complex functions, this approach is promising to outperform classical models, which often misestimate query execution costs due to simplifying assumptions.
Importantly, many \lcms rely on a graph-based representation designed to capture the properties of the physical query graph.
As shown in \cite{heinrich2025}, recent graph-based models typically achieve the best performance for cost estimation, and thus, we pick the state-of-the-art \zeroshot approach for this demo as a representative example \cite{hilprecht2022}. 
Importantly, this approach employs \gnns to learn the query execution costs from a graph representing the query plan, filter conditions and tables.
However, due to the black-box nature of \gnns, it remains unclear how the model came to its prediction, which hinders debuggability and thus troubleshooting.
Finally, our ideas can also be applied beyond \gnns, as explainers for node importance also exist for other model architectures.

\subsection{Methodology} 
The fundamental question for this work is: 
\textit{How to explain the predictions of a \lcm?}
In particular, we want to understand why the model predicts a runtime for a given plan but was highly off in its prediction.
Clearly, the predictions depend both on the trained model and the model input, which is the featurized query plan.
Thus, to answer this question, our approach provides local post-hoc explanations for a given query plan and a trained \lcm.
In particular, we focus on the \textit{node importance}, which specifies how important the presence of a given node (i.e., a query operator in a cost model) was for the prediction.
The node importance thus highlights the importance of given query operators for the runtime prediction.
For instance, if a query plan employs an expensive join operator that largely affects the runtime, this should be reflected in its node importance.
Mismatches between the operator runtimes and the provided importance scores indicate an incorrect model assumption of the \lcm about the query plan.
To quantify this, we propose novel metrics in \Cref{subsec:metrics}, which we also discuss in a more detailed example in \Cref{subsec:scenario}.
To obtain node importance scores, we adapted recent explanation algorithms (cf. \Cref{sec:algos}).

\subsection{Explanation Metrics}\label{subsec:metrics}
As a core contribution, we propose new explanation metrics relying on node importance scores that are included in our demo:
\begin{enumerate}[leftmargin=*, nosep]
    \item \noindent{}\textbf{Node Ranking:}
    \textit{Which nodes did the model prioritize the most in its predictions?}
    After obtaining the importance scores, we sorted them in descending order to identify the nodes that contributed most to the prediction.
    This ranking offers valuable insights into the model's internalized knowledge, such as which operators are consistently overlooked, poorly considered, overemphasized, or fail to align with the expectations of database experts regarding their execution costs.
    \item \noindent{}\textbf{Runtime Correlation:}
    \textit{How well does the node importance match the actual runtime?}
    In addition to the node ranking, we argue that the importance scores of given operators in the query plan should correlate with their actual runtime, i.e., longer-running operators should have a higher impact on cost estimation. 
    To report this correlation, we visualize the normalized runtime and normalized importance scores to make the ratios easily comparable.
    A mismatch of runtime fraction and importance fraction again shows false assumptions and mistakes of the \lcm.
    Note that we subtract the runtimes of the sub-plans of a given node to isolate its runtime.
    Similarly, we include the correlation with cardinalities, and we visualize correlation metrics like Spearman's rank.
    
    \item \noindent{}\textbf{Explanation Quality:} \textit{How reliable are the explanations?}
     The quality of the explanations also plays an essential role when using them for making \lcms debuggable.
     Importantly, explanations can vary depending on the explainer, and not every explanation has the same quality.
     To ensure that the explanations are reliable, we thus adapted standard evaluation metrics in the demo to assess the quality of the explanations.
     We report \texttt{Fidelity+} and \texttt{Fidelity-} \cite{yuan2023}, which describe how close a varied prediction that is based on the explanation is to the prediction of the original input graph.
     For \texttt{Fidelity+}, we compare the prediction using the query graph without the most important nodes according to the explainer and compare it against the original prediction.
     We evaluate similarly for \texttt{Fidelity-}, omitting the least important nodes.
     To apply this metric for \zeroshot models, we zeroed out the hidden states of the corresponding node to mask the node out but not affect message passing \cite{hilprecht2022}.
     Finally, we report the \texttt{CharacterizationScore} as a harmonic mean of these metrics \cite{kakkad2023}.
     In our demonstration, we scale all metrics to 0 and 1, where 1 means a higher quality.
\end{enumerate}

\begin{figure}
    \centering
    \includegraphics[width=\linewidth]{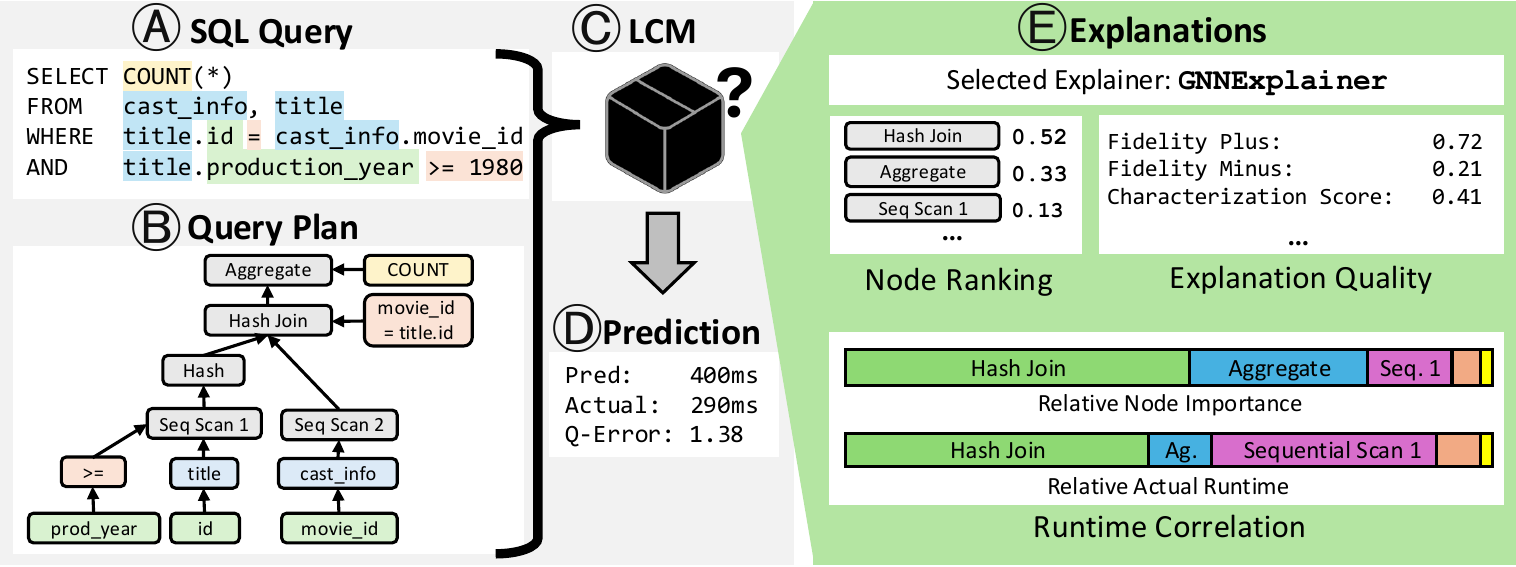}
    \caption{Example Query with a prediction and explanations.
    Execution costs for a given SQL query should be obtained \circles{A}.
    For this, it is translated to a query plan \circles{B} that employs additional nodes for filter conditions and join conditions \cite{hilprecht2022} and passed to the \lcm \circles{C}.
    The \lcm predicts query execution costs \circles{D}, which, however, can be off. 
    For this reason, our demo provides explanations of the prediction \circles{E} for a given explainer in the form of node ranking, runtime correlation and evaluation metrics.}
    \label{fig:detailed_plot}
\end{figure}

\section{Demonstration Overview}
In this section, we give an overview of our demo by presenting an example scenario and the novel explanation algorithms for \lcms, which are the main contributions of this work. 

\subsection{Example Scenario}\label{subsec:scenario}
In the following, we discuss an example scenario of our approach as shown in \Cref{fig:detailed_plot}.
This example closely follows the interaction flow of our demo that is presented in detail in \Cref{sec:demo}.
\circles{A} A SQL query that operates on the IMDB dataset is used as input for cost estimation.
\circles{B} Next, the query is converted into a query graph that contains the physical operators.
In this work, we follow the \zeroshot featurization \cite{hilprecht2022}, which, in addition to the physical operators, introduces graph nodes, e.g., for filter predicates (e.g., \predicate) or table columns (e.g., \column).
\circles{C} The physical query plan is passed to the \lcm, which uses it for cost prediction.
\circles{D} The \lcm predicts a runtime that deviates from the actual runtime. 
The prediction quality is typically expressed as Q-Error \cite{kipf2019}.
In this example, the Q-Error is 1.38, indicating a mis-prediction of $38\%$.
\circles{E} Finally, our demo allows to \textit{explain} the predictions the \lcm by providing the explanation metrics from \Cref{subsec:metrics}.

In our demo, we include the importance scores and previously introduced explanation metrics of several explainers that we extended for \lcms (cf. \Cref{sec:algos}).
For example, \gnnexplainer (which is one of our explanation algorithms) returns the following node importance scores and explanation metrics:
\textit{(1) Node Ranking}: According to the node ranking, the \hashjoin is the most important node with an importance of 0.52.
Interestingly, the \aggregate is the second most important node, having an importance score of 0.33.
\textit{(2) Runtime Correlation}: We see that the relative node importance and relative runtimes fractions have similarities but do not really match for all operators. 
While the \hashjoin is most important in both the actual runtime and the provided node importance, this is not the case for the \aggregate, which gets a higher node importance score than it should have according to its short runtime.
This indicates that the \lcm overestimated the importance of aggregations in this query plan.
\textit{(3) Explanation Quality}: Finally, we also report on the explanation quality using a metric called \texttt{Fidelity+}, which is 0.72 in the example, indicating that the explanation successfully determined the most important nodes for the prediction.
However, the least important nodes were less accurately estimated with a \texttt{Fidelity+}-Score of 0.21.
Furthermore, the \texttt{CharacterizationScore} is 0.41, indicating an average explanation quality.

\subsection{Explainer Algorithms} \label{sec:algos}
In the following, we introduce the explainer algorithms that we adapted for explaining \lcms.
Importantly, we selected explainers that apply to \gnn-based cost models, but the ideas could also be extended to other model architectures.
All these explainers are also provided in our demo.
Overall, the selected explanation algorithms provide the node importance, which is a score describing how important a given node was for a prediction.
We normalize the scores to values between 0 and 1, where a high value indicates a high importance during the prediction.
We focused on two classes of explanation algorithms: 

\noindent\textit{Gradient-based approaches} directly leverage the trained gradients of the neurons to explain the predictions of the models.
In particular, they analyze how small changes in input affect the model's output.

\noindent\textit{Perturbation-based approaches} are derived from the idea that the output of a model changes differently with respect to various input perturbations like removed nodes or altered features.
Removing an important input node or feature changes is expected to change the output significantly, while removing an unimportant node or feature should not impact the prediction.
To apply this for \zeroshot cost models \cite{hilprecht2022}, we zeroed out the hidden states of a given node to mask it out.
To adapt these approaches for regression, we introduce a relative loss when comparing the masked with the actual predictions.
We adopted the following explanation algorithms:
\begin{enumerate}[nosep, leftmargin=*]
\item \sensitivity \cite{baehrens2010} is a gradient-based explainer. 
It assumes that for a given query plan, higher absolute gradient values indicate that the corresponding features are more important than the others. 
Thus, it uses the values of the gradients and transforms them into importance scores for the input nodes.

\item \guidedbackprop \cite{springenberg2014} is also a gradient-based method that, in contrast, additionally clamps negative gradients to zero during the backward pass.
The intuition is that negative gradients are difficult to interpret, so only positive gradients are considered.
That way, \guidedbackprop guides the explanation of the features that affect the output, which is designed to improve the interpretability of explanations. 

\item \gnnexplainer \cite{ying2019} is, in contrast to the other explainers, a perturbation-based method. 
To obtain importance scores for nodes and node features, it learns soft masks for nodes to explain predictions via mask optimization. 
First, it randomly initializes soft masks for both nodes and node features and treats them as trainable variables. 
Then, an optimizer is used to minimize the difference between the original prediction and the masked prediction while maximizing the sparsity of the mask.
\gnnexplainer considers the graph structure in the explanations by design, as it is specifically designed to explain the predictions of \gnns. 
Still, it is more computationally expensive as it needs to be trained.
\item \diffmask:
Additionally, we propose \diffmask as a perturbation-based explainer.
It is based on the idea that the prediction from the masked input graph is different from the prediction from the original graph.
We derive node importance directly from the difference between the prediction with the corresponding node masked and the original prediction.
\end{enumerate}

\begin{figure}
    \centering
    \includegraphics[width=\linewidth]{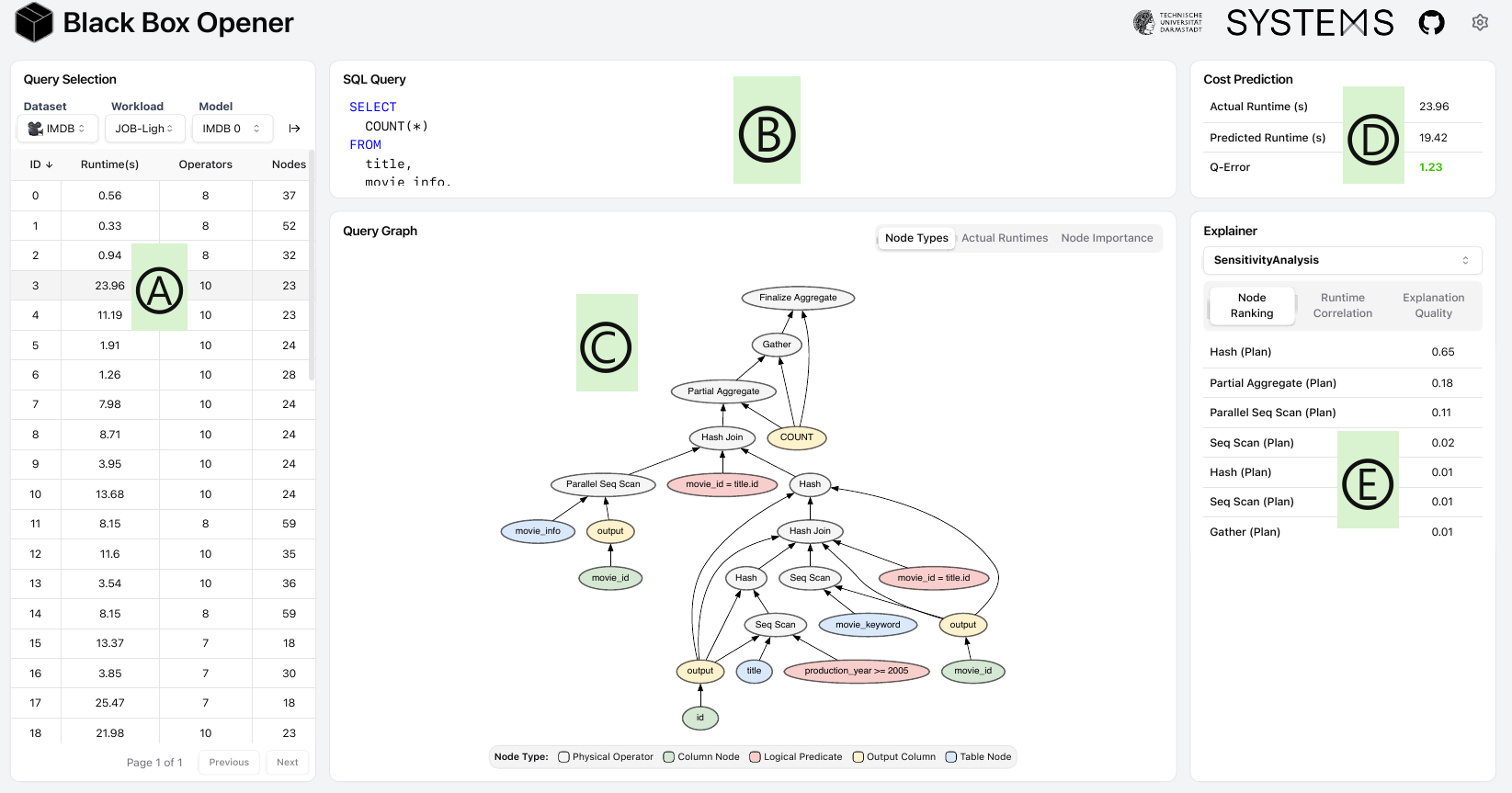}
    \caption{Overview of our Demo: \circles{A} First, a query plan is selected from given datasets and workloads.
    \circles{B} The SQL query is visualized.
    \circles{C} The query plan consists of operator, table and predicate nodes. 
    \circles{D} The predicted and actual runtime is shown together with the estimation error (Q-Error). 
    \circles{E} Finally, an explainer can be selected, and the explanation metrics are displayed.}
    \label{fig:screenshot}
    \vspace{-1ex}
\end{figure}

\subsection{User Interface}\label{sec:demo}
In the following, we describe how users can interact with our demonstration as shown in \Cref{fig:screenshot}:
Overall, our demo is an interactive web-based tool, allowing us to browse and explore datasets, query plans, models, and explanations interactively.
\circles{A} At first, the dataset is selected. 
We include several datasets from standard cost estimation datasets, including IMDB, TPC-H, and Baseball.
Moreover, queries from a selected workload can be chosen, like JOB-Light for IMDB. 
In addition, we include various synthetic workloads with varying query complexity.
For each workload, a list of query plans is shown along with various metadata, such as the number of plan operators etc., such that users can select plans of different complexity.
Once a query is selected in \circles{A}, the SQL query string is shown in \circles{B}, and in \circles{C}, users can see the query plan.
Here, all nodes and edges of the GNN-based model are displayed.
\circles{D} shows the predicted and actual runtime along with the Q-Error for the selected \lcm.
\circles{E} most importantly shows the result of our explanation. 
Here, the explainer algorithm can be selected with a drop-down bar. 
Below, the user can see the metrics discussed in \Cref{subsec:metrics}.
In addition, our demo colorizes the nodes in \circles{B} according to the importance scores or the actual runtimes to allow better comparisons.

\section{Summary and Outlook}
\noindent{}\textbf{Initial Insights.}
When using our demo to explain \zeroshot cost models, it provided interesting initial insights.
Often, we observed a high correlation between the actual runtimes and the node importance scores.
However, we also observed accurate predictions where the node importance did not really match the actual runtimes.
This shows that although the model did not always learn what it actually should, it still can be accurate.
Still, we believe this mismatch must be overcome to provide more robust predictions.
In addition, we observed that the aggregation nodes are often highly important, which contradicts their actual runtime.
This potentially shows that the model puts too much emphasis on aggregation nodes.

\noindent{}\textbf{Outlook.}
Overall, we showed that our demo provides valuable insights into the prediction behavior of \lcms and thus helps to make them explainable.
This paves the road for future improvements for \lcms, affecting model and feature design, training strategies and data collection to improve reliability.
We see further potential by exploring more dimensions, such as adding more recent explainer algorithms.
Another interesting avenue is to not only use node importance but instead evaluate feature importance.
Similarly, the scope can be extended to sub-graph importance, i.e., investigating which sub-plan of the query was most important for the prediction.

\begin{acks}
This work is funded by the LOEWE program (Reference III 5 - 519/05.00.003-(0005)), etaGPT project under grant number 03EN4107, hessian.AI and Athene Young Investigator Programme at TU Darmstadt, as well as DFKI Darmstadt.
\end{acks}

\bibliographystyle{ACM-Reference-Format}
\bibliography{bibliography.bib}
\end{document}